\documentclass[iop,apj,12pt]{emulateapj}
\usepackage{times}
\usepackage[normalem]{ulem}
\usepackage{graphicx}
\usepackage{natbib}
\usepackage{amsfonts}
\usepackage{amsmath}
\usepackage{multirow}
\usepackage{enumerate}
\usepackage{comment}
\usepackage[usenames]{xcolor}
\usepackage[plainpages=false, colorlinks=true, anchorcolor=blue, linkcolor=blue, citecolor=blue, bookmarks=false]{hyperref}
\newcommand{\be}{\begin{eqnarray}}
\newcommand{\ee}{\end{eqnarray}}
\newcommand{\lp}{\left(}
\newcommand{\rp}{\right)}
\newcommand{\lb}{\left[}
\newcommand{\rb}{\right]}

\newcommand{\shortauth}{Piro}
\newcommand{\slugcom}{Accepted for publication in The Astrophysical Journal Letters}
\slugcomment{\slugcom}

\lefthead{\sc \footnotesize \slugcom \hfill \shortauth}
\righthead{\sc \footnotesize \slugcom \hfill \shortauth}

\DeclareMathSymbol{@}{\mathord}{letters}{"3B}

\begin{document}

\normalsize


\title{Using Double-peaked Supernova Light Curves to Study Extended Material}

\author{Anthony L. Piro}

\affil{Carnegie Observatories, 813 Santa Barbara Street, Pasadena, CA 91101, USA; piro@obs.carnegiescience.edu}

\begin{abstract}
Extended material at large radii surrounding a supernova can result in a double-peaked light curve. This occurs when the material is sufficiently massive that the supernova shock continues to propagate into it and sufficiently extended that it produces a bright first peak. Such material can be the leftover, low-mass envelope of a star that has been highly stripped, the mass associated with a wind, or perhaps mass surrounding the progenitor due to some type of pre-explosion activity. I summarize the conditions necessary for such a light curve to occur, describe what can be learned about the extended material from the light curve shape, and provide an analytic model for fitting the first peak in these double-peaked supernovae. This is applied to the specific case of a Type Ic super-luminous supernova, LSQ14bdq. The mass in the extended material around this explosion's progenitor is measured to be small, $\sim0.3-0.5\,M_\odot$. The radius of this material can be $\sim500-5@000\,R_\odot$, but it is difficult to constrain due to a degeneracy between radius and the supernova's energy. In the future, spectra taken during the first peak will be important for measuring the velocity and composition of the extended material so that this degeneracy can be overcome.
\end{abstract}


\section{Introduction}

Observations of supernovae (SNe) during the first few days of their rising light curves provide valuable information about their progenitors and the circumstellar environment of the explosion \citep{Piro13}. Although historically it has been difficult to catch SNe at such early moments, current and forthcoming wide-field surveys make this an ideal time to put more focus on early light curves.

One of the exciting results from such observations is the discovery of double-peaked SNe, where the first peak lasts about a day, the second peak lasts a couple weeks, and the luminosity in each peak is relatively similar. It is important  to note that single bands at short wavelengths can also show double peaks (for example, $B$-band for SN 1987A) due to the temperature evolution of the ejecta (first heated by a shock and then later heated by radioactive nickel), and this is generally expected for a normal massive star \citep{Nakar10}. {\it This type of double-peaked light curve is distinctly different than the cases I focus on here.} The key difference is that when two peaks are seen in $R$- or $I$-band it points to more unique stellar structure \citep{Nakar14}.

Perhaps most famous among the double-peaked light curves have been the subclass of SNe IIb that exhibit this feature, such as SN 1993J, 2011dh, and 2013df \citep{Wheeler93,Arcavi11,VanDyk14}. It is now generally accepted that their first peak comes from the presence of low-mass ($\sim0.01-0.1\,M_\odot$), extended ($\sim10^{13}\,{\rm cm}$) material \citep{Woosley94,Bersten12,Nakar14}. This unique structure is also reflected in pre-explosion imaging, which identified the progenitors as yellow supergiants \citep{Aldering94,Maund11,VanDyk11,VanDyk14}.

In principle, extended material can be present around a wider variety of different SN progenitors, and thus it is useful to develop a flexible theory that can be utilized to study all of these cases. Most recently, LSQ14bdq was discovered to be a double-peaked super-luminous SN Ic where the first peak lasted $\sim15\,{\rm days}$ \citep{Nicholl15}. Although such an energetic explosion is clearly in a different regime than the previously mentioned SNe IIb, similar physics should apply. Previous work on double-peaked SNe has focused on detailed numerical models \citep{Woosley94,Bersten12}, which are difficult to use for investigating a wide parameter space, or analytic scalings \citep{Nakar14}, which do not provide time-dependent light curves for direct comparison with observations. In the present work I bridge this gap by developing a time-dependent light curve models that can be fit to the observations.

In Section \ref{sec:double peak}, I summarize the condition necessary for a double-peaked SN light curve and then present simple scalings connecting the first peak light curve to the properties of extended material, which have been utilized previously in the literature. These scaling are used in Section \ref{sec:model} to develop an analytic, time-dependent model of the first peak. In Section \ref{sec:14bdq}, I utilize this model to constrain the properties of extended material in LSQ14bdq. In Section \ref{sec:conclusion}, I summarize the conclusions and discuss future work.

\section{Conditions needed for a double peak and analytic scalings for the first peak}
\label{sec:double peak}

I begin by describing the conditions needed to produce a double-peaked SNe and summarizing the main features of their light curves. This relies heavily on the analytic results of \citet{Nakar14}, but I provide them here since they are used extensively in the next section. The basic set up is an explosion of a core with mass $M_c$ and radius $R_c$, which is surrounded by extended material with mass $M_e$ and radius $R_e$. As the SN shock powered with an energy $E_{\rm sn}$ passes from the core to the extended material, it transitions to a new velocity $v_e$. This continued propagation of the shock can only occur if $M_e$ is sufficiently optically thick such that
\be
	M_e \gtrsim \frac{4\pi R_e^2}{\kappa} \frac{c}{v_e} = 6\times10^{-5}\,\kappa_{0.34}^{-1} v_9^{-1}R_{13}^2\,M_\odot,
	\label{eq:M_e}
\ee
where the opacity is $\kappa_{0.34} = \kappa/0.34\,{\rm cm^2\,g^{-1}}$, as appropriate for electron scattering of solar composition material, $v_9 =v_e/10^9\,{\rm cm\,s^{-1}}$, and $R_{13}=R_e/10^{13}\,{\rm cm}$. The velocity in the extended material estimated considering the reverse shock that occurs at the abrupt density drop and energy conservation, resulting in
\be
	v_e \approx 2\times10^9\,E_{51}^{0.5}\lp\frac{M_c}{M_\odot}\rp^{-0.35}\lp \frac{M_e}{0.01\,M_\odot}\rp^{-0.15}{\rm cm\,s^{-1}},
	\nonumber
	\\
\ee
where $E_{51}=E_{\rm sn}/10^{51}\, {\rm erg}$. This implies that only an amount of energy
\be
	E_e \approx 4\times10^{49}\,E_{51}\lp\frac{M_c}{M_\odot}\rp^{-0.7}\lp \frac{M_e}{0.01\,M_\odot}\rp^{0.7}{\rm erg},
	\label{eq:E_e}
\ee
is passed into the extended material.

The shock-heated extended material  subsequently expands and cools, producing the first peak. The timescale for this emission to reach its maximum luminosity is set by equating the expansion and diffusion timescales \citep{Arnett82}, which is roughly estimated as
\be
 	t_p\approx \lp \frac{M_e\kappa}{v_ec}\rp^{1/2},
	\label{eq:tp}
\ee
and the peak luminosity is
\be
	L_p \approx \frac{E_eR_e}{v_et_p^2}.
	\label{eq:lp}
\ee
Equation (\ref{eq:tp}) has the correct scalings, but it is missing a prefactor that is necessary for quantitative comparisons with observations. For this reason, \citet{Nakar14} fit to numerical work \citep{Woosley94,Bersten12}, resulting in
\be
	t_p \approx 0.9\, \kappa_{0.34}^{1/2} E_{51}^{-1/4} \lp\frac{M_c}{M_\odot}\rp^{0.17} \lp\frac{M_e}{0.01\,M_\odot}\rp^{0.57} {\rm day},
	\label{eq:tp2}
\ee
and
\be
	L_p \approx 2\times10^{43}\, \kappa_{0.34}^{-1} E_{51}R_{13} \lp\frac{M_c}{M_\odot}\rp^{-0.7}
	\nonumber
	\\
	\times\lp\frac{M_e}{0.01\,M_\odot}\rp^{-0.3} {\rm erg\, s^{-1}}.
	\label{eq:lp2}
\ee
Although Equations (\ref{eq:tp2}) and (\ref{eq:lp2}) can be utilized to constrain the values of $M_e$ and $R_e$ from observations, it would also be useful to have time-dependent light curves for more detailed comparisons as well. This motivates the model presented in the next section.

\section{One-zone Model for the First Peak}
\label{sec:model}

A time-dependent light curve of the first peak is derived using a one-zone model where the entirety of the extended material is described with characteristic values. The radius of the extended material expands as
\be
	R(t) = R_e + v_et,
\ee
so that it has a characteristic density given by
\be
	\rho(t) = 3M_e/4\pi R(t)^3.
\ee
Following the first law of thermodynamics, the variation of specific entropy is related to the characteristic internal energy density $U=3E/4\pi R(t)^3$ by
\be
	TdS = \frac{1}{\rho}dU + \frac{4}{3}Ud\lp \frac{1}{\rho}\rp = \frac{1}{\rho}dU + \frac{4}{\rho}\frac{v_e}{R(t)}Udt.
\ee
The entropy decreases due to radiative cooling (I assume no radioactive heating in the extended material). Heat balance is then given by
\be
	T\frac{dS}{dt}M_e = - L
\ee
resulting in
\be
	\frac{4\pi R(t)^3}{3}\left[ \frac{dU}{dt} +\frac{v_e}{R(t)}4U\right] = - L
\ee
This expression can be written in a simpler form by substituting $E$ for $U$ and performing some algebra,
\be
	\frac{dE}{dt} + \frac{E}{t+t_e} = - L,
	\label{eq:energy}
\ee
where $t_e = R_e/v_e$ is the expansion timescale.

The radiative luminosity is approximated as
\be
	L \approx \frac{(t+t_e)E}{t_p^2}.
	\label{eq:lum}
\ee
as is expected from numerous previous analytic works \citep[e.g.,][]{Arnett82,Popov93}. With this choice of the radiative luminosity, Equation (\ref{eq:energy}) is solved analytically, resulting in
\be
	L(t) = \frac{t_eE_e}{t_p^2}\exp\lb -\frac{t(t+2t_e)}{2t_p^2}\rb.
	\label{eq:solution}
\ee
This correctly gives $L\approx t_eE_e/t_p^2$ at early times, consistent with Equation (\ref{eq:lp}). The mains limitations of this approach is that it does not include corrections to the luminosity and  temperature during the rise that are dependent on the density structure of the stellar profile. For a red supergiant structure, $L\propto t^{-0.17}$ and $T_{\rm eff}\propto t^{-0.56}$ \citep[for times later than the shock crossing time,][]{Nakar10}, while, assuming a black body spectrum, the model here gives $L\sim{\rm constant}$ and $T_{\rm eff}\propto t^{-0.5}$. Since this mostly impacts the earliest parts of the rise where data is relatively sparse, ignoring these details is a good compromise because the main point of the model is to capture the peak and fall afterwards. If necessary though, these scalings can be used to improve the solutions for $t\lesssim t_p$, and this would be especially important for bluer bands where the temperature evolution makes a larger impact. At late times, the luminosity falls as $\sim\exp(-t^2/2t_p^2)$. This solution is not accurate if recombination in the extended material begins. In all the cases used in this paper, the light curves begin to rise due to radioactive nickel heating before sufficiently low temperatures are reached, and thus recombination can be ignored.

\begin{figure}
\epsscale{1.25}
\plotone{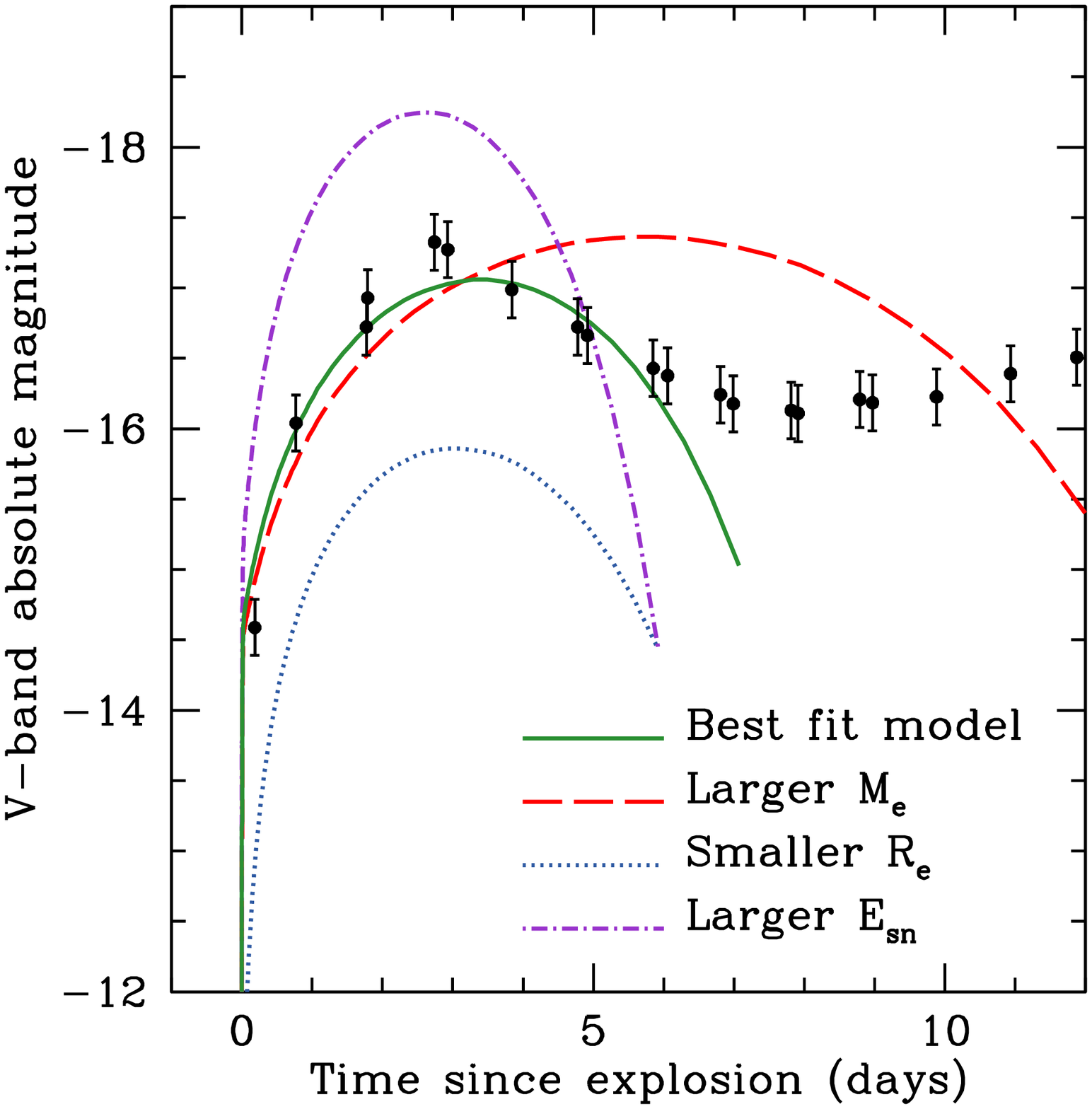}
\caption{Example first peak light curves found from Equation~(\ref{eq:solution}). Black data points are for SN 1993J \citep{Barbon95,Nicholl15}. The green solid curve is the best fit model, which uses $M_e=0.03\,M_\odot$, $M_c=2.5\,M_\odot$, $R_e=4\times10^{13}\,{\rm cm}$ and $E_{\rm sn}=0.6\times10^{51}\,{\rm erg}$. The opacity is electron scattering for solar-composition material. The other curves use all the same parameters, except in each case one parameter is changed. For the red dashed curve $M_e=0.1\,M_\odot$, for the blue dotted curve $R_e=5\times10^{12}\,{\rm cm}$, and for the purple dot-dashed curve $E_{\rm sn}=2.5\times10^{51}\,{\rm erg}$. See the text for a further discussion of the impact each of these variables have on their respective light curves.}
\label{fig:examples}
\epsscale{1.0}
\end{figure}

I provide examples of the range of light curves using Equation (\ref{eq:solution}) in Figure \ref{fig:examples}. The $V$-band absolute magnitude is calculated assuming black body emission. A cut off in the light curve is added for times less $R_e/c$ to replicate light travel effects. Data from SN 1993J are also shown as an example of a prototypical first peak from an SN IIb \citep{Wheeler93}. The green curve best fits this data, and uses values of $M_e=0.03\,M_\odot$, $M_c=2.5\,M_\odot$, $R_e=4\times10^{13}\,{\rm cm}$ and $E_{\rm sn}=0.6\times10^{51}\,{\rm erg}$. This core mass was chosen to be similar to the values typically found in \citet{Woosley94} and \citet{Bersten12}.  The other curves show examples of when one of these values is changed to provide some idea about how strongly dependent the light curve is on these parameters. In particular, $M_e$ and $E_{\rm sn}$ impact the width of the light curve, as can be seen in Equation (\ref{eq:tp2}) where these parameters have the strongest dependencies. On the other hand, $R_e$ and $E_{\rm sn}$ impact the peak luminosity. Typically $R_e$ can vary for a given stellar progenitors much more (just think about the difference between a Wolf-Rayet star and a red supergiant) than $E_{\rm sn}$ (which is usually around $\sim10^{51}\,{\rm erg}$). This is the reason light curves from shock cooling are so useful for constraining radii \citep[as emphasized in][]{Piro13}.

Another aspect to note is that the best fit extended mass of $M_e=0.03\,M_\odot$ is much less than values of $\sim0.1\,M_\odot$ that is typically presented by \citet{Woosley94} and \citet{Bersten12} for SNe IIb. This is because in fact the first peak is most sensitive to only the mass near the maximum radius and not the total amount of hydrogen present \citep[see the more detailed discussion in][and in particular their Fig.~2, which explicitly shows how the mass measured by the first peak compares with the total hydrogen shell mass]{Nakar14}. So while the total hydrogen can indeed be $\sim0.1\,M_\odot$ to produce a realistic, hydrostatic model, the first peak itself only can be utilized to measure $M_e=0.03\,M_\odot$. With this model in hand, I now turn to a very different kind of double-peaked SN to explore if this same model can be applied.

\section{Application to LSQ14bdq}
\label{sec:14bdq}

Super-luminous SNe Ic are hydrogen/helium-poor explosions that reach peak absolute magnitudes of brighter than $-21$ \citep{Quimby11,GalYam12}. Although they are relatively rare \citep[less than $\sim1\%$ of SNe,][]{Quimby13,McCrum15}, their huge luminosities make them observable at cosmological distances. Nevertheless, their underlying power source is still unclear. Unlike super-luminous SNe II, they do not show a clear signature of interaction with circumstellar material. So although this in principle is still a potentially viable model for generating their luminosity, other scenarios like spin down of a newly born magnetar \citep{Kasen10,Woosley10} and fallback accretion onto a black hole within the SN \citep{Dexter13} are also under consideration. There is considerable diversity though, and some super-luminous SNe Ic are even candidates for pair-instability SNe \citep{GalYam09,Young10} where a massive core (greater than $\sim65\,M_\odot$) undergoes complete thermonuclear disruption \citep{Heger02}.

Most recently, LSQ14bdq is a super-luminous SN Ic that was discovered to have a double-peaked light curve \citep{Nicholl15}. Previously,  SN 2006oz was another super-luminous SN with an early peak \citep[or at least an early plateau,][]{Leloudas12}, but the early detection of the rise of LSQ14bdq makes it an especially useful case for applying this theory. Although there is some uncertainty in the exact time of explosion, the first peak reaches a maximum luminosity at $\sim7\,{\rm days}$ after the explosion and a minimum occurred $\sim15\,{\rm days}$ after explosion. The brighter and broader main peak occurs $\sim50\,{\rm days}$ after explosion. \citet{Nicholl15} disfavor having the first peak powered by radioactive nickel. The problem is that its luminosity would require more nickel than the total mass needed to explain the $\sim15\,{\rm day}$ timescale width (a typical SN Ia lasts just a little longer but is considerably dimmer, showing how a radioactively-powered model is difficult to reconcile with the first peak).

\citet{Nicholl15} instead favor cooling emission of the shock-heated surface layers of the exploding star, similar to the models studied by \citet{Nakar10} and \citet{Rabinak11}. From this they infer that the radius of the exploding star must be large, $\sim500\,R_\odot$. While such emission can indeed produce a double-peaked light curve, as was observed for SN~1987A, this was in $B$-band. LSQ14bdq was instead observed in $V$-band at early times. In redder bands it is difficult to make a double-peaked light curve with the normal structure of a red or blue supergiant because as the ejecta expands and cools, it gets brighter in redder bands even if it becoming less luminous bolometrically. Therefore if the first peak is simply shock cooling it would not be as tightly peaked. So even if the radius that \citet{Nicholl15} infer may be roughly correct, the amount of mass being shocked must be much lower than what they consider. Instead it should be like the extended material I have been discussing here. This event therefore represents an exciting new opportunity to study extended material around a totally different type of SN.

\begin{figure}
\epsscale{1.25}
\plotone{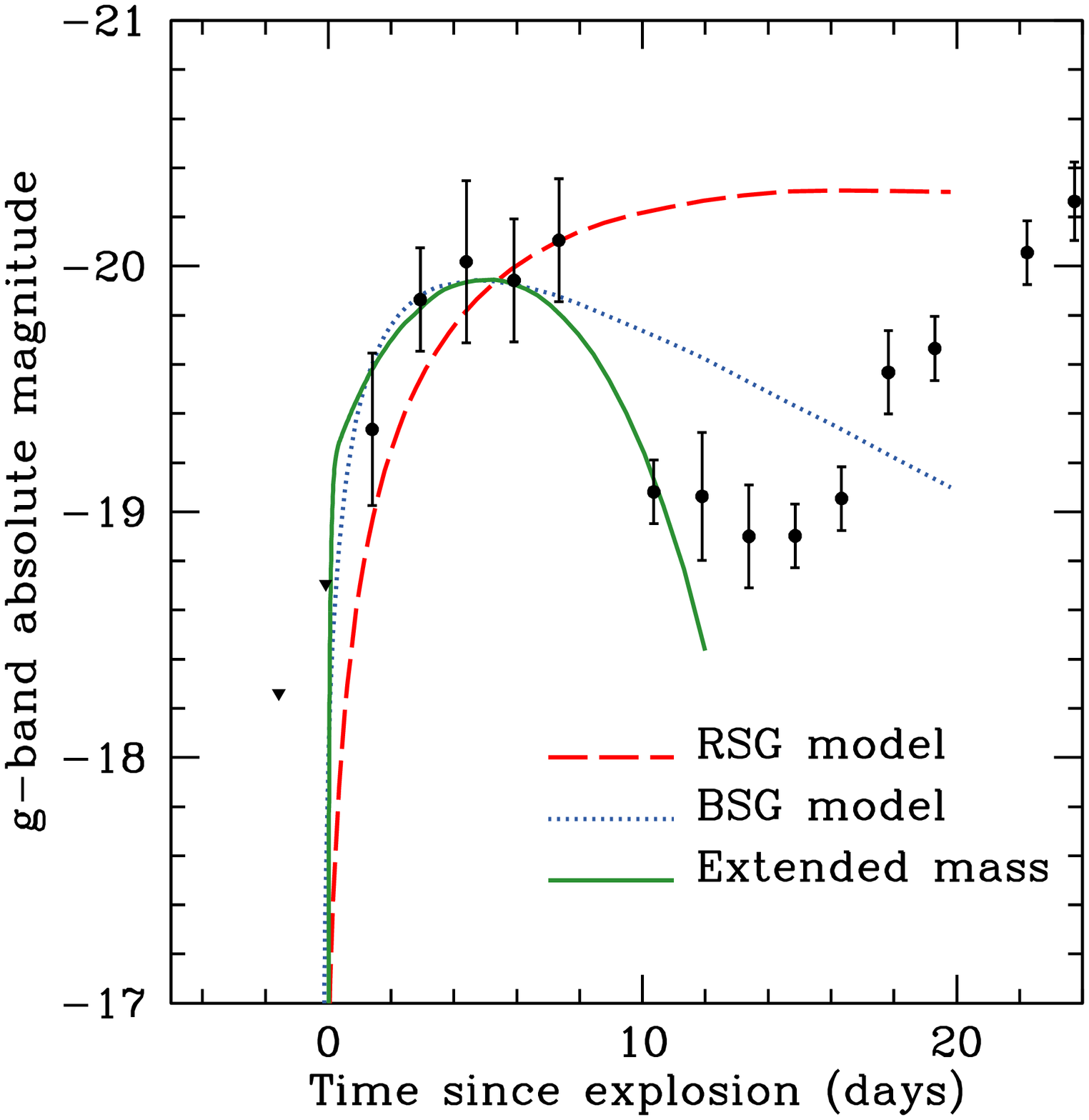}
\caption{Example fits to the first peak of LSQ14bdq using three different models. The blue and red curves are for stellar models with radii of $100\,R_\odot$ and $500\,R_\odot$, respectively, similar to those presented in \citet{Nicholl15} using the work of \citet{Rabinak11}. These roughly match the rise, but are not able to reproduce the drop that begins after $\sim7\,{\rm days}$. The green curve shows an extended material model with $M_e=0.30\,M_\odot$, $M_c=30\,M_\odot$, $R_e=5@000\,R_\odot$, and $E_{\rm sn}=8.8\times10^{51}\,{\rm erg}$.}
\label{fig:14bdq}
\epsscale{1.0}
\end{figure}

Unlike typical SNe, for which $E_{\rm sn}\sim10^{51}\,{\rm erg}$ is a resumable assumption, it is much less clear what to use for the case of a super luminous-SN. This problem is compounded because there are degeneracies between $E_{\rm sn}$, $M_e$, and $R_e$, and only $g$-band is available for LSQ14bdq for the first peak. For these reasons, I consider both the best fits to the data and then try constraining certain parameters to make additional fits. This helps highlight which values are most certain and what range is reasonable. The first comparison is in Figure \ref{fig:14bdq}, where I look at the extended model versus the shock cooling of a blue or red supergiant. These use the analytic scalings from \citet{Rabinak11} with values of $R_*=100\,R_\odot$, $M_*=60.7\,M_\odot$, $E_{\rm sn}=1.52\times10^{53}\,{\rm erg}$, and $f_\rho=0.05$ for the blue supergiant model and $R_*=500\,R_\odot$, $M_*=29.7\,M_\odot$, $E_{\rm sn}=1.78\times10^{52}\,{\rm erg}$, and $f_\rho=0.05$ for the red supergiant model ($f_\rho$ is a density parameter that only weakly impacts the light curves, see \citealp{Rabinak11} for a more detailed discussion). These specific values are chosen, and \citet{Rabinak11} is used rather than \citet{Nakar10}, to provide results similar to those presented in \citet{Nicholl15}. The main conclusion to notice is that these standard stellar models cannot provide a sufficiently quick drop in comparison to the observed first peak.

In contrast, I also plot in Figure \ref{fig:14bdq} an extended material model with $M_e=0.30\,M_\odot$, $M_c=30\,M_\odot$, $R_e=5@000\,R_\odot$, and $E_{\rm sn}=8.8\times10^{51}\,{\rm erg}$. The opacity is assumed to be $\kappa=0.2\,{\rm cm^2\,g^{-1}}$, since this event showed no signs of hydrogen, but the lack of spectra during the first peak mean that the exact opacity is uncertain. This core mass is chosen to be similar to the value \citet{Nicholl15} find by fitting the second, bright peak of LSQ14bdq with a magnetar model, and this value is used throughout the rest of this paper for the extended material models. Since $M_c$ just sets the conversion between $E_{\rm sn}$ and $E_e$ via Equation (\ref{eq:E_e}), using a different value of $M'_c$ simples implies a new energy of
\be
	E'_{\rm sn} = (M'_c/M_c)^{0.7}E_{\rm sn}.
	\label{eq:energyconversion}
\ee
Thus, for this model if I instead used $M_c=10\,M_\odot$, the corresponding energy would need be $E_{\rm sn}=4.1\times10^{51}\,{\rm erg}$ to provide a similar fit to the data. Figure \ref{fig:14bdq}  shows that this extended material model provides a much better description of the first peak, but the physical picture is now qualitatively very different than what is usually considered. Instead of just a simple massive star, there is the massive star surrounded by extended, low mass material.

\begin{figure}
\epsscale{1.25}
\plotone{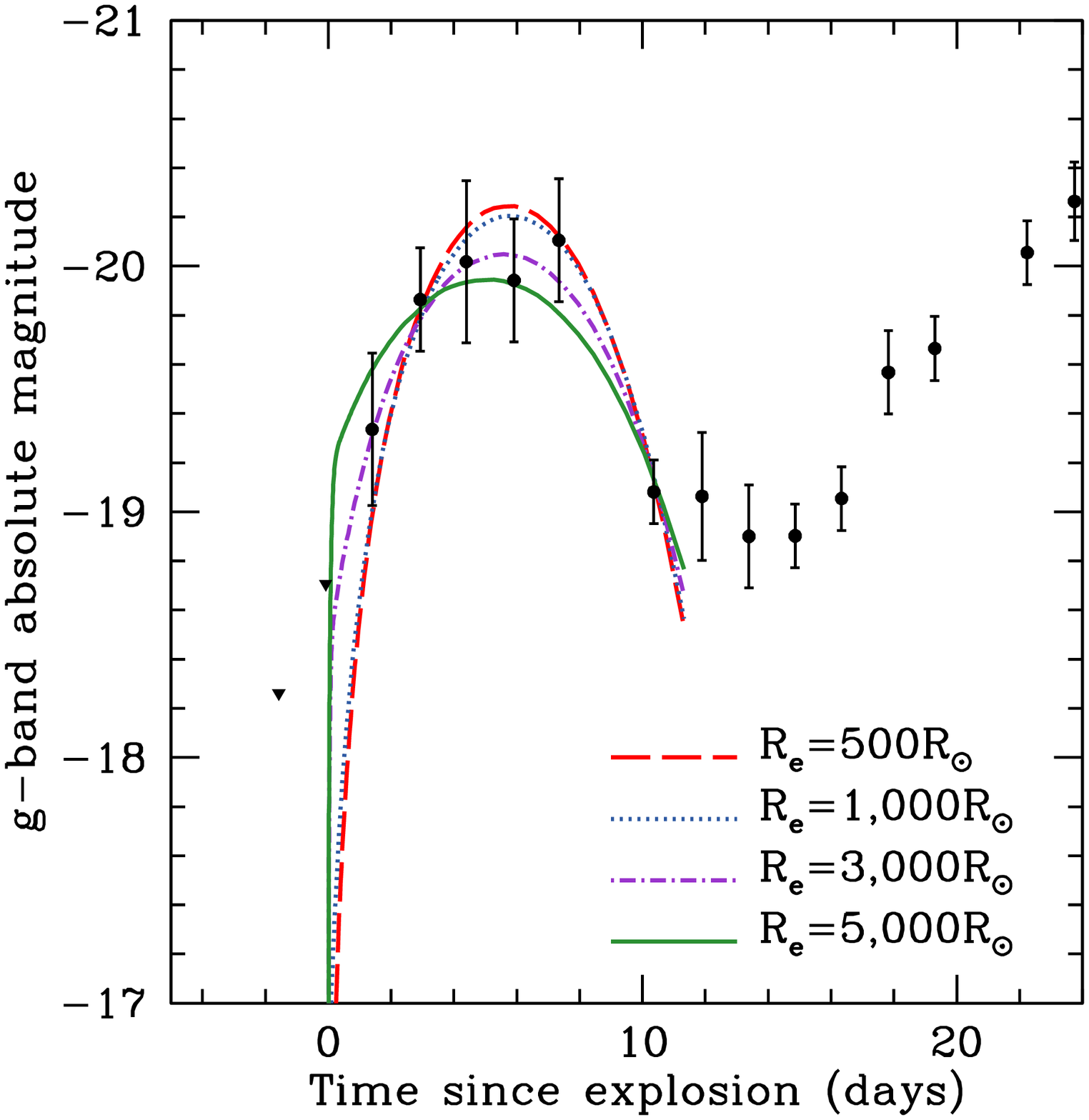}
\caption{A diverse set of extended material models that can adequately explain the first peak of LSQ14bdq. The specific values used in these models are summarized in Table \ref{tab:models}.}
\label{fig:14bdq2}
\epsscale{1.0}
\end{figure}

  \begin{deluxetable}{ccccc}
  \tablecolumns{5} \tablewidth{0pt}
  \tablecaption{Extended Material Models for LSQ14bdq}
  \tablehead{
	$M_e$ & $R_e$  & $E_{\rm sn}$  & $E_{\rm e}$  & $v_e$  \\
	 ($M_\odot$)  &  ($R_\odot$) &  ($10^{51}\,{\rm erg}$) & ($10^{51}\,{\rm erg}$)  &  (${\rm km\,s^{-1}}$) 
    }
  \startdata
	0.51 & 500 & 54.9 & 3.4 & 28,000 \\
	0.39 & 1,000 & 35.0 & 1.8 & 23,000 \\
	0.30 & 3,000 & 15.2 & 0.66 & 16,000 \\
	0.30 & 5,000 & 8.8 & 0.39  & 12,000
   \enddata
      \tablenotetext{}{All models use $M_c=30\,M_\odot$ and $\kappa=0.2\,{\rm cm^2\,g^{-1}}$. Other values for $M_c$ change the associated $E_{\rm sn}$, which can be inferred using Equation~(\ref{eq:energyconversion}).}
\label{tab:models}
\end{deluxetable}

Unfortunately, the extended model I use in Figure \ref{fig:14bdq} does not uniquely fit the observations. As can be seen from the strong dependence on both $E_{\rm sn}$ and $R_e$ in Equation (\ref{eq:lp2}), there are degeneracies between these two parameters. For more typical core-collapse SNe, assuming $E_{\rm sn}\sim10^{51}\,{\rm erg}$ would be reasonable, but this cannot be done for a super-luminous SN. To highlight how much uncertainty this introduces, in Figure \ref{fig:14bdq2} I plot a number of different solutions that can fit the data, the associated parameters of which are summarized in Table \ref{tab:models}. This shows that the best constrained parameter is $M_e$, which is set by the width of the first peak. This should be expected because the strongest dependence for $t_p$ in Equation~(\ref{eq:tp2}) is on $M_e$. This robustly shows that $M_e$ is relatively low, in the range of $\sim0.3-0.5\,M_\odot$. One note of caution is that this only represents the mass that is mostly at $R_e$. Similar to models of SNe~IIb \citep{Woosley94,Bersten12}, if this extended material is the configuration of a hydrostatic envelope, the total mass in the envelope may be greater than what is measured with $M_e$.

On the other hand, there are a wide range of values for $E_{\rm sn}$ and $R_e$ that are possible. In particular, $R_e$ needs to be large and can vary from $\sim500-5@000\,R_\odot$. The lower value is similar to \citet{Nicholl15}, but the larger values are somewhat greater than what one might consider for a standard supergiant model. Above $R_e\sim5@000\,R_\odot$ is not possible because the envelope becomes too diffuse and the shock cannot continue propagating as given by Equation (\ref{eq:M_e}). A key way to discriminate between these possible solutions would be spectra during the first peak to estimate $v_e$. A larger $v_e$ would signify a smaller $R_e$, and vice versa. I include associated values of $v_e$ in Table \ref{tab:models} for reference. Such spectra would also measure the composition of the extended material, which is also unconstrained for the models I calculate (here I simply use $\kappa=0.2\,{\rm cm^2\,g^{-1}}$, motivated by the lack of hydrogen and helium seen at later times). One might think that the fact LSQ14bdq is super-luminous, then $E_{\rm sn}$ is relatively large and (consulting Table~\ref{tab:models}) $R_e$ relatively small. Unfortunately, this does not have to be the case because the high luminosity of the second, main peak could be powered by delayed energy injection \citep[such as in the magnetar model,][]{Kasen10,Woosley10} and thus the energy associated with the first peak could be much lower like a typical SN. Again, spectra and velocity measurements during the first peak would help remove this uncertainty. Mapping the time-dependent velocity evolution better might also help probe the interior structure of the star \citep{Piro14}.

One deficiency in utilizing these theoretical light curves is that they do not constrain the distribution of mass within the extended material. This makes it difficult to discern between an extended shell that is left over from mass loss (which would have $\rho\sim r^{-3}$ or $\rho\sim r^{-3/2}$ depending on whether the material was radiative or convective, respectively, although a large radius would favor a convective profile), a stellar wind (which would have $\rho\sim r^{-2}$), or even some other non-trivial mass distribution due to pre-explosion activity of the progenitor. More detailed numerical investigations of the first peak light curve would provide more insight into how well these properties can be constrained.

\section{Conclusions and Discussion}
\label{sec:conclusion}

The discovery of double-peaked SN light curves is an exciting development that promises to provide new and unique information about the progenitors of SNe \citep{Nakar14}. LSQ14bdq has provided an opportunity where the theory for such light curves can be applied to a super-luminous SN Ic, which can help shed light on the many uncertainties that surround these relatively rare events. Motivated by this, I develop an analytic model to constrain the mass and radius associated with the first peak in this and other double-peaked SNe. This is summarized by Equation (\ref{eq:solution}), which can used along with Equations  (\ref{eq:E_e}) and (\ref{eq:tp2}) for setting the initial energy and diffusion time of the extended material, respectively. The main conclusion for LSQ14bdq is that the extended material is relatively low mass ($\sim0.3-0.5\,M_\odot$) in comparison to the bulk mass associated with the SN. The radius is more difficult to constrain, although it must be relatively large ($\sim500-5@000\,R_\odot$) to explain the high luminosity of the first peak.

In the future, continued discoveries and studies of double-peaked SNe will help inform us about the diversity of progenitors that can give rise to SNe. In most cases, the first peak will likely not be as broad as for LSQ14bdq, and thus will require cadences of less than a day to be properly resolved (as has been seen for the double-peaked SNe IIb). Short cadence surveys like Zwicky Transient Facility \citep[ZTF,][]{Law09} will be important for such studies. A wide-field UV survey \citep{Sagiv14} would be especially well-suited for this work, since the temperatures in this first peak are relatively high, $\sim10@000-20@000\,{\rm K}$. The key will be catching the first peak sufficiently early, so that the time of explosion can be constrained \citep{Piro13}. This is required to properly measure the width of the first peak, and thus the mass associated with extended material. Spectra taken during the first peak are necessary for measuring the energetics and composition of the material, so that its radius can be properly constrained. Since the range of possible velocities is large, relatively low-resolution spectroscopy will be sufficient.

Further numerical modeling should be conducted to test the applicability of these analytic results. This would be useful for exploring how efficiently energy is transferred between the core and extended material by the traversing shock, for which I used Equation (\ref{eq:E_e}) here. Such modeling would also help explore the impact of the density profile of the extended material, whether it be convective, radiative, a wind, or some other mass distribution. The recently released open source SuperNova Explosion Code \citep[SNEC,][]{Morozova15} is ideally suited for addressing these questions.

\acknowledgements
I thank Matt Nicholl for answering questions about LSQ14bdq and for helping attain data on SN 1993J. I also thank Giorgos Leloudas, Viktoriya Morozova, Ehud Nakar, and Stephen Smartt for helpful feedback and the anonymous referee for useful comments.

\bibliographystyle{apj}

\end{document}